\documentclass[conference]{IEEEtran}

\usepackage{cite}
\usepackage{dblfloatfix}
\usepackage{amsmath,amssymb,amsfonts}
\usepackage{algorithmic}
\usepackage{graphicx}
\usepackage{epstopdf}
\usepackage{siunitx}
\usepackage{bm}
\usepackage{placeins}
\usepackage{verbatim}
\usepackage{capt-of}
\usepackage{textcomp}
\usepackage{xcolor}
\usepackage[nolist]{acronym} 
\usepackage{url}
\usepackage{nidanfloat}
\usepackage{CJKutf8}
\usepackage{dashrule}
\usepackage[bottom]{footmisc} 
\usepackage{setspace}

\DeclareSIUnit \pu { pu }
\DeclareSIUnit \var { var }

\let\Phi\varPhi
\let\Theta\varTheta
\let\Psi\varPsi
\let\epsilon\varepsilon

\usepackage{pgfplots}
  \pgfplotsset{compat=newest}
  \usetikzlibrary{plotmarks}
  \usetikzlibrary{arrows.meta}
  \usepgfplotslibrary{patchplots}
\usepackage{tikz}
\usepackage{grffile}
\usepackage{todonotes}

\newcommand{\pik}{\textsuperscript{\begin{CJK*}{UTF8}{min}フ\end{CJK*}}}
\newcommand{\ub}{\textsuperscript{\begin{CJK*}{UTF8}{min}ホ\end{CJK*}}}
\newcommand{\tud}{\textsuperscript{\begin{CJK*}{UTF8}{min}ゼ\end{CJK*}}}
\newcommand{\fau}{\textsuperscript{\begin{CJK*}{UTF8}{min}ヨ\end{CJK*}}}
\newcommand{\luh}{\textsuperscript{\begin{CJK*}{UTF8}{min}マ\end{CJK*}}}
\newcommand{\tub}{\textsuperscript{\begin{CJK*}{UTF8}{min}リ\end{CJK*}}}
\newcommand{\pa}{\textsuperscript{\begin{CJK*}{UTF8}{min}パ\end{CJK*}}}

\begin{document}
\IEEEoverridecommandlockouts
\IEEEpubid{\makebox[\columnwidth]{978-1-6654-3597-0/21/\$31.00~\copyright2021 IEEE \hfill} \hspace{\columnsep}\makebox[\columnwidth]{ }}

\begin{acronym}
	\acro{DER}[DER]{distributed energy resource}
	\acroplural{DER}[DERs]{distributed energy resources}
	\acro{DG}[DG]{distributed generator}
	\acroplural{DG}[DGs]{distributed generators}
	\acro{ADN}[ADN]{active distribution network}
	\acroplural{ADN}[ADNs]{active distribution networks}
	\acro{FRT}[FRT]{fault ride through}
\end{acronym}

\title{Probabilistic Stability Assessment for Active Distribution Grids}

\author{
\IEEEauthorblockN{Sebastian Liemann\tud\textsuperscript{*}, Lia Strenge\tub, Paul Schultz\pa, Holm Hinners\ub,\\ Johannis Porst\fau, Marcel Sarstedt\luh, Frank Hellmann\pik }

\IEEEauthorblockA{
\tud Institute of Energy Systems, Energy Efficiency and Energy Economics (ie$^3$),\\ Technical University of Dortmund, Dortmund, Germany \textsuperscript{*}sebastian.liemann@tu-dortmund.de\\
\tub Control Systems Group, Technische Universität Berlin, Berlin, Germany \\
\pik\pa RD4, Potsdam Institute for Climate Impact Research (PIK), Potsdam, Germany \\
\ub Institute of Automation, University of Bremen, Bremen, Germany\\
\fau Institute of Electrical Energy Systems, Friedrich-Alexander University Erlangen-Nürnberg, Germany \\
\luh Institute of Electric Power Systems, Leibniz University Hannover, Hannover, Germany
 }
}

\maketitle
\IEEEpubidadjcol

\vspace{-2mm}
\begin{abstract}
This paper demonstrates the concept of probabilistic stability assessment on large-signal stability in the use case of short circuits in an active distribution grid. Here, the concept of survivability is applied, which extends classical stability assessments by evaluating the stability and operational limits during transients for a wide range of operating points and failures. For this purpose, a free, open-source, and computationally efficient environment (Julia) for dynamic simulation of power grids is used to demonstrate its capabilities. The model implementation is validated against established commercial software and deviations are minimal with respect to power flow and dynamic simulations. The results of a large-scale survivability analysis reveal i) a broad field of application for probabilistic stability analysis and ii) that new non-intuitive stability correlations can be obtained. Hence, the proposed method shows strong potential to efficiently conduct power system stability analysis in active distribution grids.
\end{abstract}

\section{Introduction} 
The operation of electrical grids is getting much more complex, due to the increasing coordination effort between \acp{DER} and sector coupling interfaces to the gas, heat or public communication system. In addition, grid operators of different voltage levels have to cooperate more closely in order to keep the electrical system stable. In this context, \acp{ADN} will play a crucial part in the coordination of future grids, to increase the number of \acp{DER} that can be integrated into the system. For example, \acp{ADN} can be used to provide ancillary services for higher voltage levels \cite{Mayorga2016}. Since the control system in an \ac{ADN} often allows a high degree of operational freedom, many different grid operating points are reachable by the operator \cite{Mayorga2016}. For a stable operation of the grid, the stability of these operating points have to be assessed in advance by dynamic simulations in order to initiate necessary countermeasures. In case of \acp{ADN}, large signal stability becomes especially relevant, as their controls and nonlinearites have to be considered for a correct stability assessment. However, due to the large operational envelope, classical stability assessments, e.g. short-circuit calculations or tripping of large generators, only cover a small part of these operating points and their corresponding stability regions. To counteract this shortcoming, probabilistic stability assessment methods in combination with a high-performance simulation tool can help to evaluate the stability for a large set of operating points and parameters \cite{Brzeski2016}. 

In particular, stability assessments based on a linearisation of the dynamics around a stable operating point are usually insufficient in the face of large (i.e. finite) perturbations due to the common nonlinearities and multistability of power systems. 
Hence, the extent of the corresponding basin of attraction becomes the determining quantity to assess the effect of large state deviations.
Direct methods based on energy functions \cite{Chiang2010,Vu2016} have a long tradition in transient stability analysis and are used to approximate the basin geometry. 
Geometric approaches, however, are subject to the curse of dimensionality, requiring high computational effort and hence are limited to small systems \cite{Lovasz2006,Gajduk2014}.
Alternatively, the volume of the basin \cite{Brzeski2017,Dudkowski2019} can be considered to quantify the stability of an operating point with respect to finite perturbations. 
Probabilistic stability measures such as basin stability \cite{Menck2013} identify the volume measure of the basin 
with the probability that a power system remains synchronised, given a specific distribution of finite perturbations. This extends the classic small and large signal stability assessment by not only evaluating single scenarios but rather a wide range of the state space around operating points.
The related concept of survivability \cite{Hellmann2016} extends this notion to include operational limits on transient deviations.
The advantage of these approaches is that the stability measures can be efficiently evaluated for large systems using Monte Carlo sampling \cite{Neumann1951,Evans2000}.

Probabilistic stability measures have already been successfully applied to power systems like swing equation networks \cite{Nitzbon2017,wolff2018power,Feld2019,Kim2019,Liu2019} or DC microgrids \cite{Strenge2017,Wienand2019}.  
However, an application to a more realistic grid model of e.g. \acp{ADN} is still missing. 
Therefore, this paper demonstrates the practical applicability of survivability by means of transient simulations of an \ac{ADN}.
As dynamic simulations with many variations of control parameters are very time-consuming, a simulation tool with high performance is needed. Here, the programming language Julia is explored with its package \textit{PowerDynamics.jl} \cite{pd} to perform these simulations. However, the focus in this paper is on demonstrating the concept of survivability and not on its computational speed of its implementation. Since \textit{PowerDynamics.jl} is still under development and many simulations have to be carried out, differences in model structures or calculation methods could potentially lead to large discrepancies in the simulation results, compared to commercial tools. 
Therefore, a simulation tool comparison between \textit{PowerDynamics.jl}, \textit{MATLAB Simulink} and \textit{DIgSILENT PowerFactory} is done first to show the accuracy of the \textit{PowerDynamics.jl} package and validate the model.\\
The paper is structured as follows. In Section \ref{ch-CigreModel}, a description about the structure and control of the implemented \ac{ADN} is given. In Section\,\ref{ch-model_valid}, the utilised tools are briefly described, where also their power flows and dynamic simulations results with \ac{FRT} are compared. The probabilistic stability assessment is carried out in Section\,\ref{ch-stability-analysis} and the paper closes with a conclusion and outlook in Section\,\ref{ch-conclusion}.
\paragraph*{Notation}{In this paper, capital letters like $U$ denote SI units, whereas small letters $u$ denote per units and underlined letters  $\underline{u}$ denote complex values.}  


\section{Model and active distribution grid control}\label{ch-CigreModel}

\begin{figure}[tbp]
\centering
\resizebox{0.80\linewidth}{!}
{\input{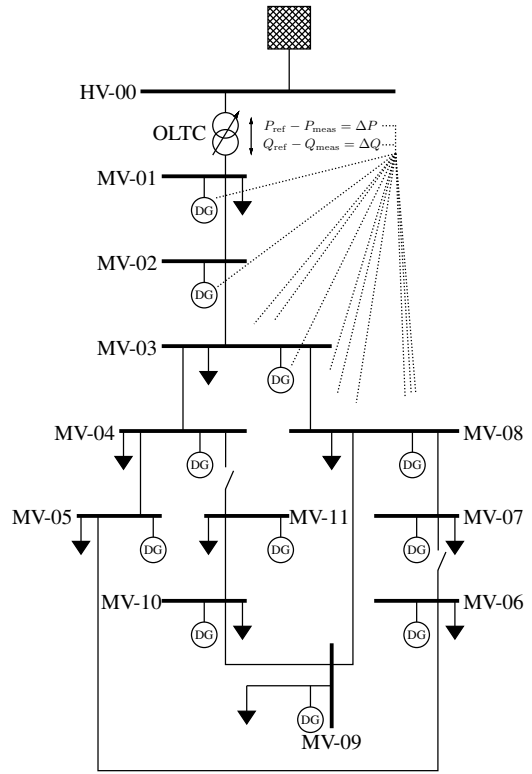}}
\caption{12-bus Cigré MV test network (own visualization, based on \cite{cigre})}
\label{cigre-full}
\end{figure}

The model of the active distribution grid is based on the 12-bus version of the CIGRE medium-voltage (MV) benchmark grid for renewable energy integration found in \cite{cigre} and shown schematically in Fig. \ref{cigre-full}. Compared to the 14-bus version it omits the second branch consisting of buses 12 to 14 and their second transformer. The parameters for the transformer, lines and loads are taken from the benchmark grid \cite{cigre}. The transformer tap is set to 10 and no phase shift is considered. 
In the following simulations, the benchmark grid is adapted such that the switches for the connection of MV-06 to MV-07 and MV-04 to MV-11 remain open and no power supply by DG units is considered during the power flow comparison (cf. Section \ref{ch-pf-comp}).
The operational limits of the distributed generators are set to $0$ and $\SI{1}{\mega\watt}$ for active power generation and $\pm\SI{1}{\mega\var}$ for reactive power generation. Depending on the simulation, the loads are either modelled as static loads without voltage dependence or as static impedance. More detailed descriptions are available in the cited sources.

\subsection{Interconnection power flow tracking}

\begin{figure*}[!htbp]
\resizebox{\linewidth}{!}
{\centering\input{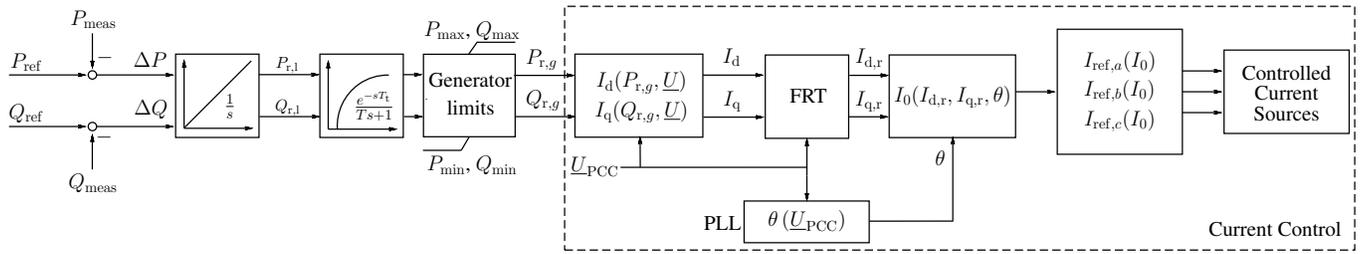}}
\caption{Distributed Generator Internal Layout}
\label{dg-internal}
\end{figure*}

In this paper, the grid is referred to as an \ac{ADN} due to its capability of following reference functions for the active and reactive power flow at the grid interconnection between the medium- and high-voltage grid. The control architecture which makes this controlled behaviour possible was developed and described in \cite{Mayorga2016}. This interconnection power flow controller is based on a distributed architecture: a \ac{DG} is connected to each bus which receives \textit{global} power flow exchange reference values $P_\mathrm{ref}$ and $Q_\mathrm{ref}$ as well as measurements $P_\mathrm{meas}$ and $Q_\mathrm{meas}$ taken at the medium-voltage side of the transformer (bus MV-01). The global control errors $\Delta P(t) = P_\mathrm{ref}(t)-P_\mathrm{meas}(t)$ and $\Delta Q(t) = Q_\mathrm{ref}(t) -Q_\mathrm{meas}(t)$ are then subjected to local integrator blocks at each medium-voltage bus which in turn yield the \textit{local} power injection reference signals $P_\mathrm{r,l}$ and $Q_\mathrm{r,l}$ determining the individual generator set points at that bus (cf. Fig.\,\ref{dg-internal}). This generator set point is then saturated using the generator limits $P_\mathrm{max}/Q_\mathrm{max}$ and $P_\mathrm{min}/Q_\mathrm{min}$. This implies that there are eleven I-controllers for active and reactive power, one at each medium-voltage bus, which all respond to the same control error measured at the interconnection. Fig.\,\ref{cigre-full} shows this signal flow using dotted lines and Fig. \ref{dg-internal} shows it for a single generator, the power flow controller being on the left of the figure. Note that the other power flow controllers receive the same $\Delta P$ and  $\Delta Q$ signals, but their integrators might yield different setpoints for each generator.

Within the \ac{DG} unit control, a $dq$-transformation is used together with a simplified model of a phase-locked loop to generate reference values for a controlled current source. An overview of the \ac{DG} control layout is illustrated in Fig. \ref{dg-internal}. The integrators are all frozen if safe voltage limits ($\pm$\SI{0.1}{\pu} is taken here) are violated at any bus and the grid then awaits stepping by the on-load tap change transformer before integration of the control error continues~\cite{Mayorga2016}.

\subsection{Fault ride-through control}
\begin{figure}[!htbp]
\resizebox{0.9\linewidth}{!}
{\centering\input{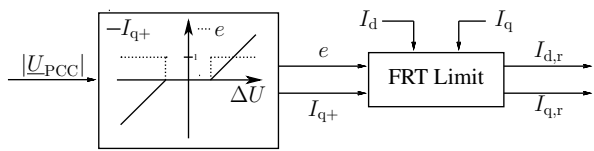}}
\caption{Details of the Fault Ride-Through control in Fig.~\ref{dg-internal}.}
\label{frt-layout}
\end{figure}

\ac{FRT} controllers are designed to support voltages in case of short-circuit events and prevent cascading loss of generation units. They accomplish this by injecting a reactive current proportional to the bus voltage error in order to support the grid voltage \cite{TARms}. In the distributed generator model described above, such a control mechanism is developed and integrated as an additional block which modifies the currents $I_\mathrm{d}$ and $I_\mathrm{q}$ depending on the voltage at the point of common coupling $\underline{U}_\mathrm{PCC}$. A schematic of the \ac{FRT} control blocks is given in Fig.\,\ref{frt-layout}. The first control block observes the magnitude of the voltage $|\underline{U}_\mathrm{PCC}|$ at the point of common coupling. The output $e=1$ represents an error flag. The dead band $U_\mathrm{dead}$ for the \ac{FRT} control is set to $\pm 10~\%$ of $\underline{U}_\mathrm{PCC}$. In such an event the block also calculates the additional reactive current injection
\begin{align}
    I_{q+}=-k_\mathrm{FRT}(|\underline{U}_\mathrm{PCC}|-U_\mathrm{ref}\pm U_\mathrm{dead}),
\end{align}
with the sign depending on whether the deviation is above or below the reference voltage (positive for negative deviation, as in line faults). The reference voltage $U_\mathrm{ref}$ is set to \SI{1}{\pu}. The block \textit{FRT Limit} then curtails $I_\mathrm{d}$ and $I_\mathrm{q}$ to the maximum generator current (i.e. $I_\mathrm{max} = \SI{1}{\pu}$) with preference given to $I_\mathrm{q}$ during voltage deviations and $I_\mathrm{d}$ otherwise.

\section{Model validation}\label{ch-model_valid}


For a comprehensive numerical analysis of the model developed in Sec.~\ref{ch-CigreModel}, the power system and \ac{ADN} control implementation in \textit{PowerDynamics.jl} is validated against established
environments like \textit{DIgSILENT PowerFactory} and \textit{MATLAB Simulink} 
with respect to the power flow solution and the short-circuit behaviour inside the dynamic simulation. Consistent results despite unavoidable tool-dependent 
differences in the model implementation indicate robustness of the  
numerical procedure.


First, \textit{PowerDynamics.jl} \cite{pd} is a package for dynamic modeling and simulation of electrical power systems written in the high-performance programming language \textit{Julia}. The power flow is calculated by determining the respective steady state of the dynamical equations using the Newton-Raphson algorithm in combination with forward-mode automatic differentiation to determine the Jacobian \cite{nlsolve}. Numerical RMS simulations are performed with the Rodas4 solver, a 4th-order stiff-aware Rosenbrock method for the RMS simulations provided by \textit{DifferentialEquations.jl} \cite{Rackauckas2017}.

Second, \textit{DIgSILENT PowerFactory} is a commercial power system analysis software that offers various models and functions for power system analysis depending on the implemented study case. 
For the load flow solution, the Newton-Raphson algorithm based on power equations are chosen, while the RMS simulations are solved with a trapezoidal method \cite{PF2020}.

Finally, \textit{MATLAB Simulink} is a programming environment for modeling, simulating and analyzing multi-domain dynamical systems. 
To model and simulate electrical power systems, the \textit{Simscape Power Systems toolbox} is used. 
Power flow calculations are based on the Newton-Raphson algorithm in \textit{power\_loadflow}.m. Dynamic investigations within the RMS domain are performed using the trapezoidal Heun method integrated in the \textit{MATLAB} suite. 

\subsection{Power flow comparison}\label{ch-pf-comp}
The initial values of the state variables within dynamic simulations are based on the steady state of the system. In electric power systems, the steady state is defined by the complex bus voltage vector $\underline{\mathbf{\mathit{U}}}_{PF}$ which is the solution of the power flow equation determined by a power flow calculation (cf. \cite{OswaldLF}). 
Differences in the power flow solution between different environments
may arise even from subtle differences in the modelling of the components used. 
A comparison of the simulation tools regarding the ADN power flow results 
reveals that the absolute deviations of the bus voltages are less than \SI{e-4}{\pu} for the magnitudes and \SI{e-3}{\degree} for the phase angles. 
Only the phase angle results in \textit{Simulink} differ slightly from the other simulation tools while an implementation in pure \textit{MATLAB} based on \cite{OswaldLF} does not show this deviation. In this comparison, the \ac{DG} units do not inject any power and only the loads are active. 
For the next sections, this power flow is used as the starting point for the subsequent dynamic simulations. Here, the measured power flow over the transformer with $P_\mathrm{meas} = \SI{24.373}{\mega\watt}$ and $Q_\mathrm{meas}  = \SI{6.115}{\mega\var}$ are used as the base reference points $P_\mathrm{ref}^0$ and $Q_\mathrm{ref}^0$.

\begin{figure}[!thb]
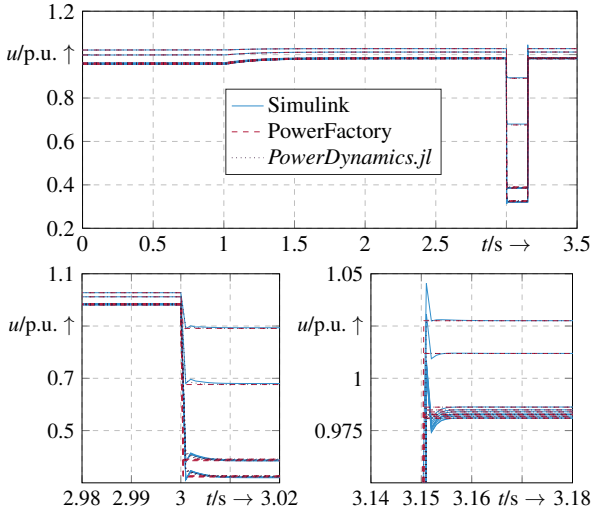

\resizebox{0.90\linewidth}{!}
{\centering\large{\input{bilder/frt-simulink1.tex}}}
\resizebox{0.90\linewidth}{!}
{\centering\large{\hspace{0.5mm}\input{bilder/frt-simulink2_new.tex}}}
\caption{Comparison of \ac{ADN} Control and \ac{FRT} behaviour in Simulink, PowerFactory and \textit{PowerDynamics.jl} for all bus voltages}
\label{frt-simulink}
\end{figure}

\subsection{Dynamic simulation and short-circuit comparison}\label{ch-frt}
The agreement in the  power flow solution extends to the behaviour of the ADN control during short-circuit events.
The dynamic simulation results of each simulation tool are plotted in Fig. \ref{frt-simulink}, where all bus voltages (except HV-00) of the test system are illustrated. The short-circuit takes place at bus\,8 at $t=\SI{3}{\second}$ with a fault-impedance of $R_\mathrm{on}=\SI{3.5}{\ohm}$ and is cleared at $t=\SI{3.15}{\second}$. Before the short-circuit, the \ac{ADN} is subjected to a reference step at $t=\SI{1}{\second}$ from  $P_\mathrm{ref}^0$ and $Q_\mathrm{ref}^0$ to $P_\mathrm{ref}=\SI{24}{\mega\watt}$ and $Q_\mathrm{ref}=\SI{5}{\mega\var}$ at the low-voltage side of the transformer to demonstrate the capabilities of the control architecture. As it can be seen by Fig. \ref{frt-simulink} the deviations during the dynamic simulations are very small between all simulation tools. Only in \textit{MATLAB Simulink}, small differences can be detected during the occurrence and disappearance of the short circuit, which can be attributed to numerical inaccuracies of the solver at these discontinuous events. In general, the results are consistent and demonstrate the applicability of the implementation in Julia for dynamic ADN simulations. Therefore, the following probabilistic stability analysis is carried out in \textit{PowerDynamics.jl} due to its computational efficiency.

\section{Probabilistic stability analysis}\label{ch-stability-analysis}
\begin{figure}[!hb]
\resizebox{0.95\linewidth}{!}
{\centering\large{
%
\begin{tikzpicture}

\begin{axis}[%
width=3.673in,
height=1.613in,
at={(0.616in,0.218in)},
scale only axis,
xmin=0,
xmax=4.25,
xtick={0.15,2.75,3.375,4},
xticklabels={{0.15},{3},{${\text{\it{}t}}\text{/s}\rightarrow$},{60}},
ymin=0,
ymax=1,
ytick={0.15,0.5,0.8,0.9},
yticklabels={{0.15},{${\text{\it{}u}}\text{/p.u.}\uparrow$},{0.85},{0.90}},
axis background/.style={fill=white},
xmajorgrids,
ymajorgrids,
grid style={dashed}
]

\addplot[area legend, draw=none, fill=green, fill opacity=0.1, forget plot]
table[row sep=crcr] {%
x	y\\
0	0.15\\
0.15	0.15\\
2.75	0.8\\
4	0.8\\
4	0.9\\
4.25	0.9\\
4.25	1\\
0	1\\
}--cycle;

\addplot[area legend, draw=none, fill=red, fill opacity=0.1, forget plot]
table[row sep=crcr] {%
x	y\\
0	0.15\\
0.15	0.15\\
2.75	0.8\\
4	0.8\\
4	0.9\\
4.25	0.9\\
4.25	0\\
0	0\\
}--cycle;
\addplot [color=red, line width=2.0pt, forget plot]
  table[row sep=crcr]{%
0	0.15\\
0.15	0.15\\
2.75	0.8\\
4	0.8\\
4	0.9\\
4.25	0.9\\
};
\end{axis}

\begin{axis}[%
width=4.74in,
height=1.979in,
at={(0in,0in)},
scale only axis,
xmin=0,
xmax=1,
ymin=0,
ymax=1,
axis line style={draw=none},
ticks=none,
axis x line*=bottom,
axis y line*=left
]
\node[below right, align=left]
at (rel axis cs:0.2,0.73) {connected\\$\text{("}{\text{\it{}survive}}\text{")}$};
\node[below right, align=left]
at (rel axis cs:0.38,0.45) { disconnected\\$\text{("}{\text{\it{}not survive}}\text{")}$};
\end{axis}
\end{tikzpicture}%
}}
    \caption{Low voltage fault-ride through limiting curve (own visualization, based on \cite{TARms})}
    \label{fig:FRT_curve}
\end{figure}
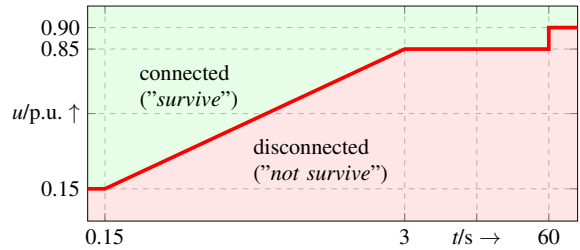

In this section, a probabilistic stability assessment, in particular the survivability, of the dynamic \ac{ADN} model is performed with respect to short-circuit clearance.\footnote{The code can be found at  \url{github.com/strangeli/adn-survivability} .}
It can be shown that this recent concept from  dynamical systems theory yields complementary information about the stability of an \ac{ADN}'s operating point.
In contrast to classical stability concepts such as the critical clearing time of generators, the concept of survivability extends analyses by covering a wide range of stability regions due to the large examination space. Especially for the survivability, the monitoring of operational limits of assets during transients are also included to further improve the assessment of the overall system stability. Thus, simulations which are numerically stable but would not occur in reality, e.g. due to protection devices, can be better evaluated in the stability assessment. To evaluate the short circuit's stability impact at one of the buses,  the $\textit{VDE-AR-N 4110}$ \ac{FRT} limiting curve \cite{TARms} shown in Fig.~\ref{fig:FRT_curve} is relied on. In particular, \textit{survival} is defined as a successful fault-ride through according to the limiting curve (cp. Fig \ref{fig:FRT_curve} and \cite{TARms}).

The fault model considered here is an ensemble of short-circuits at medium-voltage buses with a random fault duration and fault resistance.
The fault duration is taken from a normal distribution $\Delta t_\mathrm{fault} = \mathcal{N}(\SI{150}{\milli\second}, \SI{10}{\milli\second})$ with a mean of $\SI{150}{\milli\second}$ and a standard deviation of $\SI{10}{\milli\second}$.
The fault resistance is randomly drawn from a uniform distribution $R_{on} = \mathcal{U}(\SI{3}{\ohm}, \SI{10}{\ohm})$.
Such a fault leads to a transient voltage drop as it is visualized in Fig.~\ref{frt-simulink}. 

In the next subsections, firstly the survivability of the whole system is estimated for short-circuits at single nodes to demonstrate its general concept. Secondly, the focus is set so the \ac{ADN} control where a large envelope of possible operating points for $P_\mathrm{ref}$ and $Q_\mathrm{ref}$ are assessed according to their short-circuit survivability. In particular, the second study shows how the survivability in combination with an efficient simulation tool helps to investigate broad stability regions using large-signal stability simulations. Here, the focus is only on the identification of safe operating regions and not how they are interpreted or used by the grid operator.

\subsection{Single-node survivability}\label{sb-surv}
The concept of {single-node survivability} \cite{Hellmann2016} is an example for a probabilistic stability assessment of deterministic systems.
In this paper, it is defined as the conditional  probability of survival at \textit{all} buses given a \textit{local} fault at a single bus according to the setup above. 
Following the approach in \cite{Hellmann2016}, the probability of bus $i$ to survive a random fault is the single-node survivability value $\mu(i\vert R_{on}, \Delta t_\mathrm{fault})$ which is estimated by using Monte-Carlo sampling.
For this, each bus of Fig.\,\ref{cigre-full} is subjected to a sample of $N = 1000$ short-circuits from the fault model and the cases $T_i$
that survive according to our definition are counted.
Since the survival of each simulation run is a binary outcome, a Bernoulli experiment is performed with empirical mean $\mu(i\vert R_{on}, \Delta t_\mathrm{fault}) = \tfrac{T_i}{N}$ and confidence 
interval bounded by $1/(2 \sqrt N) \approx 1.6\%$ \cite{agresti1998approximate} 
independent of the state space dimension.
\begin{figure}
    \centering
    \includegraphics[width=0.90\columnwidth]{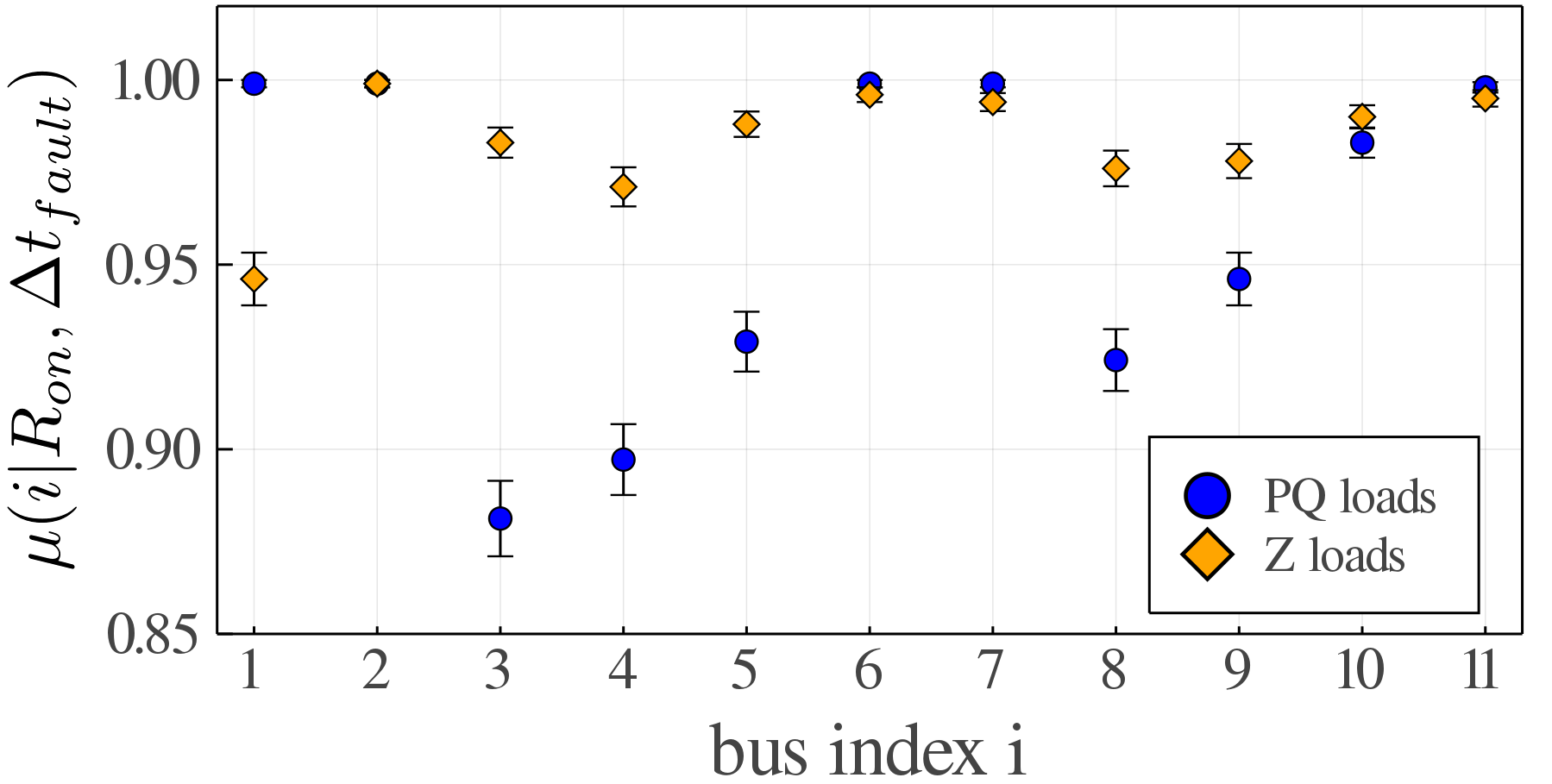}
    \caption{Single-node survivability}
    \label{fig:single-node-surv}
\end{figure}

Fig. \ref{fig:single-node-surv} shows the estimated single-node survivability for short circuits at the respective bus on the \textit{x}-axis for voltage-independent (\textit{PQ}) and static impedance (\textit{Z}) loads. 
It is impacted both by the load model and the location of the fault.
The results show that bus 3 has the lowest survivability in case of constant loads, which is not obvious, as it is more centrally connected compared to other nodes which is usually beneficial in case of short-circuits. However, a reason might be the high line impedance between buses 2 and 3. As the short-circuit at bus 3 leads to low voltages at the buses behind, the loads are drawing high currents to keep their power constant. All these currents have to flow over the line 2-3 which results into an even bigger voltage drop and thus decreasing the probability of a successful low-voltage ride through. 
At the same time, the DG units are not capable of lifting the fault voltage further within their FRT limits, which could have reduced the flow between buses 2 and 3.
This also explains the high survivability for buses at the end of the feeder (e.g. bus 6 and 11), as only at these buses the voltage is low and the voltage drop over the line. For constant impedance loads, this effect cannot be observed, as the load current decreases.  
Overall, the modeling of loads has a strong impact on the survivability in view of the different results. This suggests that a survivability analysis can discover non-intuitive effects, but can also confirm known relationships in terms of stability. 

\subsection{$P_\mathrm{ref}$ and $Q_\mathrm{ref}$ envelope survivability}

\begin{figure}[tb]
    \includegraphics[width = 0.95 \columnwidth]{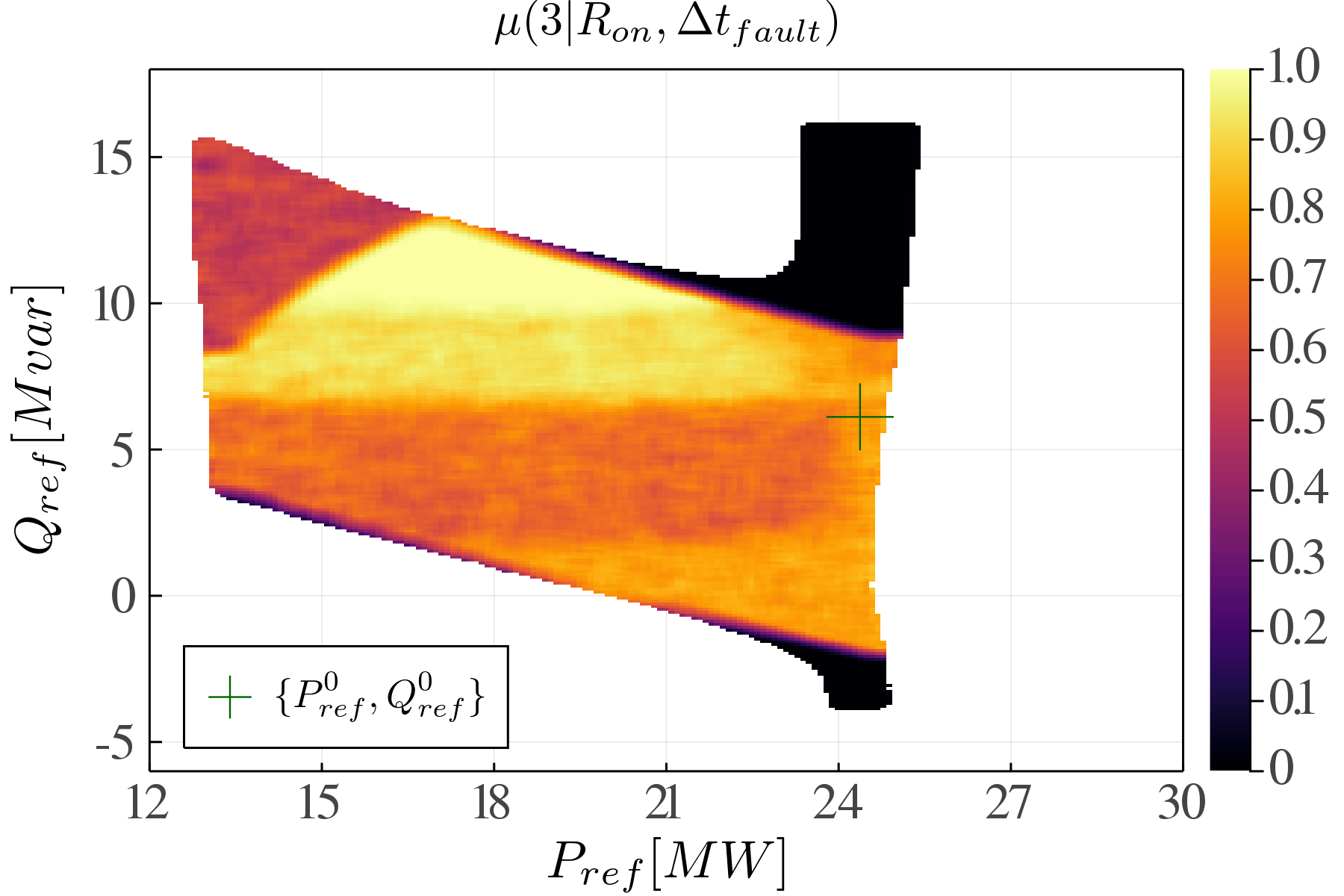}
    
\vspace{-8pt}\centering{ \hdashrule{0.85\linewidth}{0.5pt}{5pt} }\vspace{4pt}%

 \includegraphics[width=0.95\columnwidth]{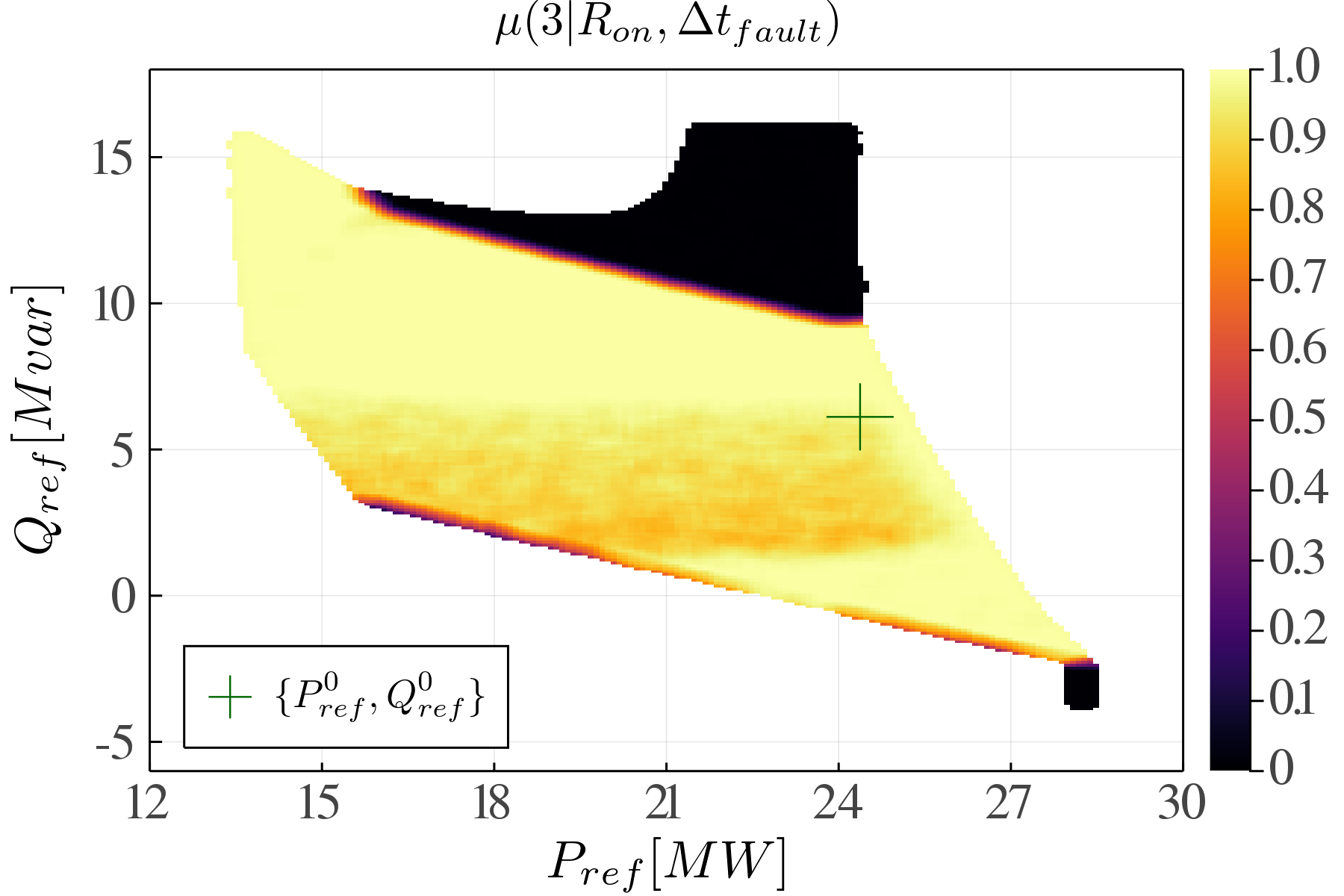}
    \caption{Short-circuit survivability at bus MV-03 for static loads (top) and for voltage-dependent loads (bottom)}
    \label{fig:surv_0}
\end{figure}

From Fig. \ref{fig:single-node-surv}, it is shown that bus 3 has the lowest single-node survivability with $\mu(3\vert R_{on}, \Delta t_\mathrm{fault})\approx 0.881 \pm 0.016$ (\textit{PQ}), defining the worst-case outcome.
Therefore, it is chosen for a more detailed analysis. In particular, the flexibility options available to the grid operator with respect to changes in the reference set points $P_\mathrm{ref}$ and $Q_\mathrm{ref}$ are of interest.
To do this, the single-node survivability analysis at bus\,3 is combined with a step change in the reference signal as in Fig.~\ref{frt-simulink}. After each step, a short circuit at bus\,3 is simulated for those samples where the relative global control error $\Delta P / P_{meas}$ is below 5\%. This numerical experiment is carried out with $10^6$ samples for $P_\mathrm{ref}$ and $Q_\mathrm{ref}$, demonstrating the computational efficiency provided by Julia which allows such large simulations to be performed at manageable cost. The output of a single simulation is a boolean value for its survivability.
Fig.~\ref{fig:surv_0} shows the voltage survivability for a random short circuit after a reference step at $t = 0.25$  \SI{}{\second} starting from the former power flow $\{P_\mathrm{ref}^0,Q_\mathrm{ref}^0\}$ to a random value taken from a uniform distribution of $P_\mathrm{ref} \in  [P_\mathrm{ref}^0 - \SI{15}{\mega\watt}; P_\mathrm{ref}^0 + \SI{15}{\mega\watt}]$ and $Q_\mathrm{ref} \in  [Q_\mathrm{ref}^0- \SI{10}{\mega\var}; Q_\mathrm{ref}^0+ \SI{10}{\mega\var}]$. The survivability is calculated by a clustering of the sampled region into overlapping squared cells of size $\SI{0.5}{\mega\watt}\times \SI{0.5}{\mega\var}$. For each cell, the survivability is calculated by the number of surviving samples -- according to the definition and fault model in Sec.~\ref{sb-surv} -- divided by the total number of samples in the cell (which is ensured to be greater or equal 100).  

Hence, the results in the Fig.~\ref{fig:surv_0} show the feasibility region of the reference steps for a fixed OLTC tap position and the variation of the single-node survivability at bus\,3 within this region for voltage-independent (Fig.~\ref{fig:surv_0}, top) and voltage-dependent loads (Fig.~\ref{fig:surv_0}, bottom). The black area of $\mu(3\vert R_{on}, \Delta t_\mathrm{fault})\approx 0$ can be reached with a reference step, but in a fault-case the FRT limits are easily violated. This is plausible since the DG units have to provide a lot of inductive reactive power and therefore the voltage in the grid decreases. With increasing active power infeed by the DG units, the voltage rises, increasing the survivability in turn. Comparing the figures, voltage-dependent loads have a larger feasibility region than voltage-independent loads and they have in general higher survivability within the feasibility region. This can be explained by the reduced power requirement of voltage-dependent loads during the fault and therefore they help maintain the voltage.  New ``forbidden'' zones arise for voltage-independent loads (upper left) and the boundaries between low and high survivability are very steep near the edges. These boundaries identified here may be worth studying in more detail in future work. 

\section{Conclusions}\label{ch-conclusion}
In this paper, the safe operating point region with respect to fault tolerance, a task expected to be of high importance to future ADN operators, is investigated. 
Concretely, it could be demonstrated that probabilistic methods can be an effective tool to handle increasing complexity in grid operation.
Such methods are highly adaptive to specific research questions and can be used iteratively for exploring stability regions and their relation to system characteristics.
Here, the study based on short-circuit survivability reveals numerous open questions, especially regarding the origin of the stability thresholds. 
Future research should target alternative failure models and parametrisations, e.g. parameter studies for limits on the short-circuit. 
The proposed method and environment also allow for an extension to other probabilistic stability measures  \cite{Leng2016,Lindner2018,Schultz2018} and for a comparison of various network-local perturbations \cite{Nitzbon2017} to gain a further deeper understanding of the stability properties of ADNs.
Remarkably enough, it could also be shown that high performance open source implementations like \textit{PowerDynamics.jl} are sufficiently capable and accurate to correctly capture the systems behaviour, making comprehensive transient stability studies accessible. However, detailed statements about the computational efficiency need to be adressed in future research.


\section*{Acknowledgment}
{\scriptsize\setstretch{0}
This work was funded by the Deutsche Forschungsgemeinschaft (DFG, German Research Foundation) – KU 837/39-1 / RA 516/13-1, project numbers 360248793, 360290054, 360460668, 359921210.
The authors acknowledge the support of BMBF, Condynet2 FK. 03EK3055A.
All authors gratefully acknowledge the European Regional Development Fund (ERDF), the German Federal Ministry of Education and Research and the Land Brandenburg for supporting this project by providing resources on the high performance computer system at the Potsdam Institute for Climate Impact Research.\par
}

\bibliography{main}
\bibliographystyle{IEEEtran}

\end{document}